\documentclass{article}
\usepackage{spconf,amsmath,graphicx}
\usepackage{xcolor}
\usepackage{multirow}
\usepackage{enumitem}
\usepackage{booktabs}
\usepackage[normalem]{ulem}
\usepackage[square,comma,numbers]{natbib}
\usepackage{hyperref}
\usepackage{float}

\title{Infant Crying Detection in Real-World Environments}
\name{Xuewen Yao$^1$, Megan Micheletti$^2$, Mckensey Johnson$^2$, Edison Thomaz$^1$, Kaya de Barbaro$^2$}
\address{
  $^1$Department of Electrical and Computer Engineering, University of Texas at Austin, USA\\
  $^2$Department of Psychology, University of Texas at Austin, USA}
\begin{document}
\ninept
\maketitle
\begin{abstract}

Most existing cry detection models have been tested with data collected in controlled settings. Thus, the extent to which they generalize to noisy and lived environments is unclear. In this paper, we evaluate several established machine learning approaches including a model leveraging both deep spectrum and acoustic features. This model was able to recognize crying events with F1 score 0.613 (Precision: 0.672, Recall: 0.552), showing improved external validity over existing methods at cry detection in everyday real-world settings. As part of our evaluation, we collect and annotate a novel dataset of infant crying compiled from over 780 hours of labeled real-world audio data, captured via recorders worn by infants in their homes, which we make publicly available. Our findings confirm that a cry detection model trained on in-lab data underperforms when presented with real-world data (in-lab test F1: 0.656, real-world test F1: 0.236), highlighting the value of our new dataset and model.


\end{abstract}
\section{Introduction}
\label{sec:intro}

Infant crying is a critical evolutionary signal that allows infants to communicate hunger, discomfort and pain. Crying is also a known stressor that can decrease caregiving quality  and increase risks for infant development and caregiver mental health \cite{meg_day}. Developing automatic cry detection algorithms that perform robustly in everyday home settings is essential for basic science investigation of the links between crying, caregiving, and caregiver mental health. Furthermore, such algorithms would be key to the development of applications to provide “just-in-time” support to caregivers.

Recently, a number of infant cry classification algorithms have been published. However, many of these were developed and evaluated using data in controlled settings \citep{infant_cry_background_2018, infant_cry_background_2017}. Others are recorded in real-world settings but are trained and evaluated on short, pre-parsed segments containing non-overlapping individual sounds \cite{crying_solved}. By contrast, real-world environments, such as family households, typically include a variety of complex overlapping sounds that must be distinguished from behaviors of interest. These additional sounds greatly increase the difficulty of detection and classification problems; additionally, if such sounds are not present in the training data, performance will deteriorate in real-world conditions. The distinct challenges of detection and classification in real-world settings relative to clean-lab conditions have been illustrated in other domains. For example, Liaqat et al. \cite{coughwatch} and Gillick et al. \cite{laughter_detection} demonstrated that cough and laugh detection models performed poorly when trained and tested on real-world datasets relative to those trained and tested on in-lab datasets. In particular, the F1 score of cough detection dropped by 13.5\% and the F1 score of laughter detection drops 20.7\% percent when trained and tested on real-world vs. in-lab data.


%
%

To attempt to mimic real-world conditions in training data, some papers have manually added sounds, such as from cars or medical equipment, to crying datasets collected in laboratory settings \citep{infant_cry_cnn_log_mel_2018}. However, as demonstrated in other domains, models tested on synthetic datasets typically yield higher accuracy than those tested on real-world datasets \cite{vad}. For example, the AUC of voice-activity detection trained and tested on a synthetic dataset was 5.27\% higher than results of a model trained and tested on a real-world dataset. Thus, synthetic datasets that layer additional sounds on clean laboratory datasets are not equivalent to real-world datasets. 


Researchers such as developmental psychologists have also relied on LENA (Language ENvironment Analysis), a commercial product, to capture and process relevant acoustic events - including infant crying - from continuous recordings collected in children’s everyday environments \cite{alex_lena}. However, the main focus of LENA is to detect parent and child speech vocalizations. While LENA has never released accuracy statistics on its cry detection models, recent work shows that its infant crying predictions have low accuracy relative to trained human coders \cite{MegICIS}. Comparing LENA against other tools and approaches is challenging since its training corpus, algorithms, and models are not made publicly available.

In summary, there is a lack of robust models for detecting infant crying from real-world home settings. Additionally, there is a lack of high-quality datasets that can be used to address this problem. Thus, in the current paper:

\begin{itemize}
  \item  We collect and annotate 780 hours of raw audio data from recorders worn by infants in home settings, including 7.9 hours of annotated crying. We make this large dataset publicly available \footnote{ \url{https://homebank.talkbank.org/access/Secure/deBarbaroCry-protect/deBarbaroCry.html}}.
  \item We evaluate multiple modeling strategies for cry detection. Our best performing model detects crying from continuous real-world environments, reaching an F1 score 0.613. We share this model on Github \footnote{\url{https://github.com/AgnesMayYao/Infant-Crying-Detection}}.
  \item To showcase the importance of training cry detection models with real-world datasets, we compare the performance of our best  model with a model trained on in-lab data and tested on real-world data. Our model achieves an F1 score of 0.656 when tested on in-lab data but drops to F1 of 0.236 when tested on real-world data.
  
\end{itemize}


\section{Related Work}
Recent reviews of infant crying detection/classification indicate that many published models achieve F1 scores over 0.9 \cite{crying_solved}. However, we argue that these models are unlikely to perform robustly in truly real-world settings. First, most supposed “real-world” training data does not reflect the actual complexity of true real-world data. For example, some real-world cry datasets were collected in home environments but only recorded discrete samples of crying rather than labelling crying from continuous audio recordings \cite{crying_manikanta}. Negative examples may also be non-representative. For example, some “real-world” models appear include discrete samples of fans, music, speech, dogs barking, or doors closing  \citep{crying_gu}. Such sampled data greatly simplifies the problem given that real-world crying - and its negative examples -  typically co-occur with other everyday sounds. Other datasets appear to be drawn from continuous real-world audio data, but are described so generally that the specifics of the training data are uncertain \cite{crying_lavner}, for example, whether all recorded audio is included or only those sounds labelled with a single identifiable source. Overall, given that these datasets are not publicly available it is impossible to assess their ecological validity and therefore the generalizability of their models in real-world conditions.

Models trained on truly real-world data have much lower accuracy. For example, using continuous infant-worn recordings, models submitted for the 2019 Interspeech Baby Sound Challenge \cite{Interspeech2019} achieved baseline unweighted average recall of up to 0.54 in multi-class classification of infant vocalizations. However, cry classification was poor, with a F1 of 0.26 (as calculated from the published confusion matrix \cite{Interspeech2019}). These results showcase the challenge of real-world cry detection and highlight that this problem is not yet solved.

Finally, all previously published cry detection models simplify the task by trimming audio recordings to cry episodes and classifying individual segmented vocalizations \citep{ Interspeech2019} rather than detecting infant crying from continuous audio, as we propose to do. 

Notwithstanding these concerns, prior work provides some insight into effective approaches for cry detection \cite{crying_solved}. Most approaches use machine learning models with acoustic input features including fundamental frequency (F0) and mel-frequency cepstrum coefficients (MFCCs) \citep{infant_cry_car_2012} and convolutional neural network models (CNNs) \citep{infant_cry_cnn_log_mel_2018}. In direct comparisons, CNNs yield better results than classic ML approaches\citep{infant_cry_cnn_vs_2020}. Additionally, recent work indicates that CNNs outperform recurrent neural networks (RNNs) in cry detection \cite{infant_cry_cnn_vs_2020}.  In the current paper, we use AlexNet\cite{alexnet}, which performs reasonably well in audio classification tasks relative to more recent models, including VGG19 and ResNet-50 \citep{spectrumSnore_2017, alexnet_better}. Additionally, it is more suitable for real-world applications and mobile computing given its relatively low power consumption and memory utilization \cite{alexnet_resources}. While spectral representations of sound have not yet been applied to cry detection, these features have been successfully used to classify other complex tonal vocalizations, such as emotion from speech \cite{spectrumEmotion_2017}. In a comparison model, we leverage deep representational features generated from the higher layers of AlexNet and compare its performance with end-to-end AlexNet.

\section{Datasets}

To develop a robust model for detecting infant crying outside of the lab, we collect infant-worn audio recordings in real-world settings and annotate crying in two subsets of these recordings, RW-Filt and RW-24h. RW-Filt contains audio recordings with high-likelihood of crying episodes. RW-24h consists of randomly sampled audio clips, which we use as a second, more rigorous testing dataset. Finally, we test how models trained with in-lab data perform on real-world data. For this evaluation, we compile an in-lab crying dataset, IL-CRIED. See Table~\ref{tab:data} for a summary of the three datasets, which are further described in the next section. 

\subsection{Real-world audio datasets}

We collected 742 hours of real-world infant-worn audio recordings as part of a broader study examining the dynamics of mother-infant activity in everyday home settings \cite{Holding_2019}. Data was collected using the LENA audio recording device, a lightweight, commercially-available audio recorder. Parents were instructed to place the LENA in a vest worn by the infant and record up to 72 hours total of audio data in their home, including two weeknights and a weekend. Thus our real-world audio recordings contained complex overlapping sounds present in natural home environments including the sounds of daily activities, vocalizations from all family members including the target child, noise from the child's clothes rubbing the sensor, as well as stretches of silence. The LENA can record up to 24 hours of audio on a single charge. Audio data was stored in PCM format with one 16-bit channel at a 22.05kHz sampling rate \cite{lenaBattery}. 

\subsubsection{Real world: Filtered Dataset (RW-Filt)} 
Given the sheer size of our recorded audio data and the imbalanced nature of infant crying in real life, prior to annotation we filtered a subset of our audio recordings to include only segments of audio that had a high-likelihood of containing crying. Filtering the data in this way allowed us to annotate positive crying examples much more efficiently than annotating by listening to continuous unfiltered audio recordings. To filter, we used algorithms provided with the LENA software to provide predictions for all infant non-speech sounds, which is a category that includes infant crying, fussing and vegetative sounds (e.g. burping). All infant non-speech sounds detected by LENA that occurred within five minutes of one another were grouped together as a single episode. If an episode contained at least 2 seconds of contiguous crying according to the LENA predictions, the entire episode was included in the filtered dataset and annotated by trained research assistants as described in Section~\ref{sec:annotation}.
 
\subsubsection{Real world: Unfiltered 24h Dataset (RW-24h)}
While our filtered dataset provided a relatively balanced and high-density source of infant crying episodes, the filtering step likely excluded other examples of crying or sounds that may be mistaken for crying. Thus, given that our goal is to create a model that can robustly detect crying from continuous real-world household audio, we also annotated a second subset of unfiltered, randomly sampled audio data to be used exclusively for testing model accuracy. While this unfiltered dataset is highly unbalanced (see Table~\ref{tab:data}), it is more representative of infant's everyday audio space and therefore a stronger standard of assessment. Our unfiltered dataset contained 17 continuous raw 24-hour audio recordings from 17 individual infants, distinct from the infants included in RW-Filt.

\subsection{In-lab audio dataset (IL-CRIED)}
To assess the importance of real-world training data on model performance, we trained an in-lab model using the CRIED database \cite{CRIED} made available as part of the Interspeech 2018 Crying Challenge \cite{interspeech2018}. The CRIED corpus was collected from microphones positioned over awake infants lying in a cot in a quiet room with no external stimuli other than occasional speech by parents. The corpus comprises 5587 individual vocalisations from 140 recordings of 20 healthy infants along with the start and end time of each vocalization within the recording. Vocalizations include infant neutral/positive, fussing, crying, and overlapping adult vocalizations. Using the start and end times provided by the authors, we re-created the original recordings by inserting silence between consecutive infant vocalizations. This was necessary because although labels at the vocalization level are technically more precise, clinicians and researchers, as well as parents, analyze and report crying at a more molar level, i.e. at the level of crying episodes. As such, a trained research assistant re-labelled the CRIED dataset with the same annotation scheme used for the RW-Filt and RW-24h datasets. We refer to the relabeled CRIED dataset as IL-CRIED. Using the same annotation scheme across both the RW and IL datasets also allowed us to do a more direct comparison of models trained with these data. Prior to using the IL-CRIED data for training, we also applied white noise to the entire dataset to provide a more realistic in-lab environment.

\subsection{Cry episode annotation (RW-Filt, RW-24h, IL-CRIED)} \label{sec:annotation}
A team of trained research assistants in our lab labelled both real-world and in-lab datasets for infant crying. Similar to existing coding schemes \cite{crying_annotation}, crying episodes included both fussing and crying vocalizations. Additionally, all cry episodes had a minimum duration of 3 seconds and were combined if neighboring episodes were within 5 seconds of each other. All other infant sounds, as well as all unlabeled household audio and silence, were collapsed into a second category labelled “not crying”. Annotations were conducted according to best practices in behavioral sciences (inter-rater reliability kappa score \cite{kappa}: 0.8469, corresponding to strong agreement).

 \begin{table}[!ht]
  \caption{Crying Dataset Statistics}
  \label{tab:data}
  \centering
  \begin{tabular}{ccccc}
    \toprule
     \textbf{Dataset} & \textbf{Cry Hrs} & \textbf{Total Hrs} & \textbf{N} & \textbf{Ages (months)} \\
    \midrule
    RW-Filt  & 7.9 & 66 & 24 & 1.53 - 10.8\\
    RW-24h & 14.7 & 408 & 17 & 0.78 - 7.03 \\
    IL-CRIED & 1.26 & 14 & 20 & 1 - 4 \\
    \bottomrule
  \end{tabular}
\end{table}

\subsection{Data preprocessing}
To create clean representations of “crying” in our training data, we segmented audio recordings into 5 second windows (with 4-second overlap). All windows containing more than one label were removed. Thus, each 5 second training segment is labelled with either “crying” or “not crying”. To balance our dataset, we first augmented crying data with a time masking deformation technique, in which we randomly masked up to 0.44 seconds in each duplicated crying segment, as described in \cite{specAugment_2019}. Next, we randomly removed “not crying” segments until the number of not crying segments matched the number of augmented “crying” windows.

To facilitate effective detection in our testing dataset, we preprocessed our audio data. Specifically, given that the mean F0 of infant crying ranges from 441.8 to 502.9 Hz \cite{cryFrequency_2003} we removed all audio segments that were silent above a 350 Hz threshold. To reduce fragmenting, we concatenated neighboring segments occurring within 5 seconds of one another and removed isolated segments shorter than 5 seconds. Smoothed data were cut into 5-second windows (with 4-second overlap) as input for testing our model.



\section{Cry Detection Models}
To explore the performance of different models on infant crying detection, we implement and evaluate two baseline machine learning architectures: a baseline SVM classifier with acoustic features (AF) and a baseline end-to-end CNN model (CNN). Additionally we test a \textit{new approach} combining a SVM classifier with deep spectrum features and acoustic features (DSF + AF). Finally, we compare the performance of our \textit{new approach} with a transfer learning method leveraging a deeper pre-trained model, VGGish \cite{CNN_audioset}. To evaluate our four models, windowed cry predictions were converted to second-by-second bins. The prediction for each second was labelled “crying” if any window containing that second had a prediction of “crying”. Additionally, to reduce fragmentation, we smoothed the predictions at the second level by eliminating “crying” and “not crying” episodes that were shorter than or equal to 5 seconds and reassigning them to the surrounding class.

\begin{table*}[!ht]
  \caption{Infant cry detection performance on both real-world and in-lab dataset, with second-by-second accuracy averaged across participants. }
  \label{tab:results}
  \centering
  \begin{tabular}{p{0.2\textwidth}cccccc}
    \toprule
    & \multicolumn{3}{c}{\textbf{Results on RW-Filt (LOPO)}}  & \multicolumn{3}{c}{\textbf{Results on RW-24h }}  \\\cmidrule(l{0.5em}r{0.5em}){2-4} \cmidrule(l{0.5em}r{0.5em}){5-7}
      \textbf{Train on RW-Filt}  & \multicolumn{1}{c}{F1} & \multicolumn{1}{c}{Precision} & \multicolumn{1}{c}{Recall} & \multicolumn{1}{c}{F1} & \multicolumn{1}{c}{Precision} & \multicolumn{1}{c}{Recall}\\
    AF& 0.515($\pm$0.185) & 0.42($\pm$0.225) & 0.847($\pm$0.140) & 0.502($\pm$0.204) & 0.481($\pm$0.239) & 0.586($\pm$0.191)   \\

    CNN & 0.620($\pm$0.182) & 0.505($\pm$0.206) & 0.873($\pm$0.110) & 0.589($\pm$0.194) &	0.642($\pm$0.217) & 0.580($\pm$0.178) \\
     
     DSF + AF & 0.615($\pm$0.170) & 0.521($\pm$0.191) & 0.820($\pm$0.147) & 0.613($\pm$0.184) & 0.672($\pm$0.219) & 0.552($\pm$0.178)\\
     VGGish & 0.574($\pm$0.204) & 	0.445($\pm$0.216) &	0.936($\pm$0.062) & 0.543($\pm$0.204) & 	0.489($\pm$0.228) & 0.652($\pm$0.182)\\
    \midrule
    \multirow{1}{*}{\textbf{Train on IL-CRIED}} & \multicolumn{3}{c}{\textbf{Results on IL-CRIED (LOPO)}}  & \multicolumn{3}{c}{\textbf{Results on RW-24h }}  \\\cmidrule(l{0.5em}r{0.5em}){2-4} \cmidrule(l{0.5em}r{0.5em}){5-7}
    DSF + AF & 0.656($\pm$0.191) & 0.578($\pm$0.255) & 0.808($\pm$0.128) & 0.236($\pm$0.122) & 0.143($\pm$0.084) & 0.851($\pm$0.162) \\
    \bottomrule
  \end{tabular}
\end{table*}

\subsection{SVM with acoustic features (AF)}
\label{sec:af}
As a baseline, we extract acoustic features and fed them into an SVM classifier for training. As frequency domain features (such as MFCCs) and time domain features (such as zero-crossing rate, short-time energy) are frequently used in infant crying detection and classification \cite{crying_solved}, we leveraged pyAudioAnalysis \cite{pyaudioanalysis} and 34 acoustic features, including zero-crosing rate, energy, and 13-element MFCCs, were extracted. Features were extracted at each second within the window and the mean, median and standard deviation of those features were calculated and those 102 features were fed into an SVM classifer with RBF kernel for training.

\subsection{End-to-end CNN model (CNN)}
\label{sec:cnn}
We used a modified AlexNet as our baseline end-to-end CNN model with mel-scaled spectrograms as input. To extract spectrograms, we applied short-time Fourier transfrom (samples per segment = 980, overlap = 490, Hann window) to each window and acquired mel-scaled spectrograms of size $225\times 225\times 1$. To match this shape, we changed the dimension of AlexNet input layer to $225\times 225\times 1$ and output layer to 2 for our binary classification problem. Additionally, we added batch normalization layers \cite{batch} after every convolutional and fully-connected layer except the output layer. Modified AlexNet was trained using Adam Optimizer \cite{kingma2017adam} (learning rate = 0.001, $\beta_1$ = 0.9, $\beta_2$ = 0.999) for 50 epochs at a batch size of 128.


\subsection{SVM with deep spectrum and acoustic features (DSF + AF)}
Next, we combined our two baseline approaches to build a new model leveraging both deep spectrum features and acoustic features. The last hidden layer with size $1000$ of the CNN model described in Section~\ref{sec:cnn} was extracted as deep spectrum features and concatenated with acoustic features (extracted as in Section~\ref{sec:af}) and fed into an SVM classifier with RBF kernel for training. 

\subsection{SVM with VGGish embeddings (VGGish)}
For our last model, we obtained embeddings of dimension $128\times 1$, extracted by VGGish, and fed them into an SVM classifier with RBF kernel for training. During the process, the pre-trained VGGish model was not fine-tuned and only the SVM classifier was trained.

\section{Results and Discussion}
We report results in Table~\ref{tab:results}. We first evaluate our models' ability to detect crying in real-world setting using the RW-Filt dataset as training data. The end-to-end CNN model achieves the highest F1 score at 0.620 using leave-one-participant-out (LOPO) cross-validation. When applied to the RW-24h dataset, the DSF + AF model achieved the highest F1 score at 0.613. As the RW-24h dataset is a completely unfiltered audio dataset, it poses a higher challenge to the classifiers. 

Examining the DSF + AF approach with lab data, the model achieved a LOPO F1 score of 0.656 when trained and tested with IL-CRIED. On the other hand, performance was considerably lower when this model was trained with in-lab data but evaluated with the RW-24h real-world dataset (F1 score: 0.236).

\subsection{Impact of model architecture on performance}
We trained and compared four different models for detecting crying. Our results indicate that the model combining deep spectrum and acoustic features achieved the best results in our 24h real-world environments. However, the combination of acoustic and deep spectrum features showed only a slight boost in performance over the end-to-end CNN model. Thus, it appears that the supervised CNN training contributed most substantially to the DSF + AF model's performance. Deep spectrum features are known to be robust to the effects of environmental noise \cite{spectrumEmotion_2017}. The VGGish model did not outperform our baseline CNN model, suggesting a relatively shallower network, when trained with enough targeted data, can attain better performance than when using embeddings from a deeper network.


\subsection{Real world vs. In-lab training data} 

To confirm if the performance of the model built using in-lab data generalizes to real-world settings, we first established a baseline with in-lab data. By training and testing the model with in-lab data, it achieved an F1 score of 0.656. While this score is lower than results presented in recent work \cite{crying_solved}, our inclusion of overlapping and adult vocalizations likely increased the difficulty of the task. Additionally, we note that this result is comparable to the Interspeech 2018 Crying Challenge baseline \cite{interspeech2018} which achieved an F1 score of 0.688 on this same dataset, using a majority voting of multiple models. 

Next, we explored how models trained with in-lab data, performed on the real-world dataset. In this case, the F1 score measure dropped to 0.335 when our the model was tested on RW-Filt and 0.236 when tested on RW-24h. We noted that models trained with in-lab data have high recall but low precision when tested on real-world datasets. This observation, i.e., high-rate of false positives, confirms that the model is not capable of discriminating crying from other acoustic events, a common difficulty across domains and applications (e.g., laugh detection \cite{laughter_detection}) when attempting to generalize classification from in-lab conditions to real-world environments.

Importantly, when our models trained on the real-world dataset, they achieved much better performance when tested on real-world data. This result underscored that robust recognition of real-world crying might be possible when a representative training dataset is used, and suggests that datasets collected in controlled environments do not represent the full complexity of real-world environments and are of limited use in the context of the crying detection task. This is important since most existing cry detection models thus far have been tested with data collected in controlled settings \citep{infant_cry_background_2017, infant_cry_background_2018}.

\section{Conclusion}

In this paper, we examined the problem of infant cry detection in real-world environments. We evaluated several machine learning strategies for classification using both in-lab and real-world data, and proposed a promising modeling approach leveraging both deep spectrum features and acoustic features. In support of this work, we collected and annotated a novel dataset of infant crying in home environments, possibly the largest audio corpus of annotated real-world crying available. Our proposed model performed significantly better in real-world environments when trained with real-world data, suggesting that crying events captured in controlled laboratory conditions have limited utility in modeling these acoustic episodes.

\bibliographystyle{IEEEbib}
\bibliography{refs}





\end{document}